\documentclass[pdflatex,sn-nature]{sn-jnl}
\usepackage{graphicx}
\usepackage{amsmath,amssymb}
\usepackage{booktabs}
\usepackage{siunitx}
\usepackage{float}
\usepackage{gensymb}
\begin{document}
\title[Article Title]{Mimicking the earth's core conditions with ultrafast laser materials interaction}
\author*[1]{\fnm{Mohamed Yaseen} \sur{Noor}}\email{noor.48@osu.edu}
\author[2]{\fnm{Aram} \sur{Yedigaryan}} 
\author[1,3]{\fnm{Gabriel} \sur{Calderon}}\email{calderonortiz.1@osu.edu}
\author[2]{\fnm{Arshak} \sur{Tsaturyan}}
\author[2]{\fnm{Elena} \sur{Kanchan}}\email{elena.kachan@univ-st-etienne.fr}
\author[1,3]{\fnm{Jinwoo} \sur{Hwang}}\email{hwang.458@osu.edu}
\author[4,5]{\fnm{Carmen S.} \sur{Menoni}}\email{Carmen.Menoni@colostate.edu}
\author[2]{\fnm{Jean-Philippe} \sur{Colombier}}\email{jean.philippe.colombier@univ-st-etienne.fr}
\author*[1,6,7]{\fnm{Enam} \sur{Chowdhury}}\email{chowdhury.24@osu.edu}

\affil[1]{\orgdiv{Department of Materials Science and Engineering}, \orgname{The Ohio State University}, \orgaddress{\street{140W 19th Avenue}, \city{Columbus}, \postcode{43210}, \state{Ohio}, \country{United States}}}
\affil[2]{\orgdiv{CNRS, Institut d'Optique Grduate School, Laboratoire Hubert Curien UMR 5516}, \orgname{Université Jean Monnet}, \city{Saint-Etienne}, \postcode{42023}, \country{France}}
\affil[3]{\orgdiv{Center for Electron Microscopy and Analysis}, \orgaddress{\street{1305 Kinnear Road}, \city{Columbus}, \postcode{610101}, \state{Ohio}, \country{United States}}}
\affil[4]{\orgdiv{Electrical and Computer Engineering Department}, \orgname{Colorado State University},\orgaddress{\street{400 Isotope Dr}, \city{Fort Collins}, \postcode{80523}, \state{Colorado}, \country{United States}}}
\affil[5]{\orgdiv{XUV Lasers Inc},  \city{Fort Collins}, \postcode{80525}, \state{Colorado}, \country{United States}}
\affil[6]{\orgdiv{Department of Electrical and Computer and Engineering}, \orgname{The Ohio State University}, \orgaddress{\street{2015 Neil Ave.}, \city{Columbus}, \postcode{43210}, \state{Ohio}, \country{United States}}}
\affil[7]{\orgdiv{Department of Physics}, \orgname{The Ohio State University}, \orgaddress{\street{191 W Woodruff Ave.}, \city{Columbus}, \postcode{43210}, \state{Ohio}, \country{United States}}}

\vspace{30mm}
\abstract{Ultrafast lasers create extreme, non-equilibrium thermodynamic conditions that can transiently reach pressures and temperatures comparable to Earth’s interior core. Here we show that femtosecond excitation of amorphous SiO$_2$/HfO$_2$ multilayer dielectrics drives the formation of high-pressure crystalline phases of silica— including stishovite, seifertite, and the pyrite-type high density structure—within confined subsurface regions. Using TEM, SAED, and 4D-STEM, we directly map nanoscale phase evolution and identify crystalline motifs embedded inside laser-generated blisters. Complementary molecular dynamics simualtions reveal the thermodynamic pathway underlying these transformations: rapid electronic pressure initiates densification and octahedral coordination, followed by temperature-driven crystallization and displacive transitions during ultrafast quenching. The resulting polymorphs reflects a dual-stage pathway inaccessible under equilibrium processing. Our results establish femtosecond laser excitation as a viable route to synthesize and stabilize ultrahigh-density high pressure silica phases under ambient conditions, without a diamond anvil cell, with implications for laser-damage mechanisms, high-energy-density materials, and planetary physics.}


\maketitle
\vspace{-7mm}
\section{Introduction}\label{sec1}
Silica (SiO$_2$), one of the most technologically significant dielectric materials, exhibits a diverse landscape of structural polymorphs that emerge under varying thermodynamic conditions. At ambient pressure and temperature, silica is characterized by tetrahedral SiO$_4$ coordination in crystalline phases such as quartz or cristobalite, or in its ubiquitous amorphous form \cite{Hemley2018}. Under extreme pressure, however, silica undergoes structural transitions to higher-density configurations. Among these, stishovite featuring octahedrally coordinated silicon is the most well-known high-pressure polymorph \cite{Salleo2003,Kresse1998,Tsuchida1990,Dubrovinsky1997}. So far, these phase transitions, which are fundamental to modeling planetary interiors, unfortunately can only be accessed temporarily within a diamond anvil cell (DAC). Moreover, they are also of growing importance in high-power laser and optical applications where silica is a key dielectric material. High-power laser facilities—spanning inertial confinement fusion, ultrafast optics, and extreme-field science—routinely drive silica into far-from-equilibrium conditions where intense, femtosecond-scale energy deposition produces transient pressures and temperatures comparable to geophysical shock events \cite{Paisner1994,Danson2019}. Despite silica’s reputation as a high bandgap, robust material, repeated or high-fluence laser exposure leads to irreversible modifications including densification, defect accumulation, and the emergence of high-pressure crystalline motifs \cite{Clnar2020,Shieh2004,Apostolova2021UltrafastReview}. For example, Salleo et al., \cite{Salleo2003} observed that nanosecond irradiation at 355 nm partially converts fused silica into defective stishovite-like material, confirmed via infrared spectroscopy and electron diffraction.

Static high-pressure experiments using DACs have significantly advanced our understanding of silica polymorphism \cite{Kuwayama2005,Luo2004,Oganov2005}. Kuwayama et al., \cite{Kuwayama2005} demonstrated the temporary formation of a pyrite-type phase characterized by 6+2 silicon coordination above 260 GPa and simultaneously heated to 1800 K using a continuous wave (CW) laser, suggesting the existence of even denser phases beyond stishovite. The interplay between the temperature, pressure range and its direction, leading to this phase transition under different conditions, is still a matter of ongoing theoretical and experimental investigations \cite{pan, pang2022newOccurrence, park2025melting}. While such pressures are traditionally accessed via static or shock compression techniques, recent advances in ultrafast laser excitation suggest that comparable conditions can be generated transiently. Femtosecond laser pulses deposit energy into the electrons inside the material on sub-picosecond timescales, which then transfer the energy into the lattice within picoseconds, leading to highly localized, non-equilibrium pressure-temperature (P–T) environments capable of driving phase transformations without bulk heating \cite{Sundaram2002}. Several studies have demonstrated the ability of ultrafast lasers to induce structural rearrangements in disordered systems \cite{Shieh2004,Werner2019,Bhuyan2017UltrafastMicroexplosion}. For instance, Shieh et al. \cite{Shieh2004} showed that amorphous silicon can crystallize under near-infrared femtosecond pulses, forming nanograins with dimensions comparable to the laser wavelength. Although silica differs substantially in bonding and optical absorption, these studies underscore the broader capability of ultrafast lasers to drive phase transitions through nonthermal pathways. In wide-bandgap materials such as silica—where linear absorption is negligible at infrared wavelengths—energy deposition is governed by nonlinear absorption and multiphoton ionization, generating dense electron–hole plasmas that can transiently modify interatomic potentials and promote atomic rearrangement and metastable phase formation.

These subtle structural transformations precede catastrophic material removal and alter local absorption, refractive index, and stress evolution, effectively lowering the laser-induced damage threshold and accelerating failure. However, the thermodynamic trajectory linking sub-critical modification to irreversible damage in silica remains unresolved. Identifying the pressure–temperature pathways, coordination changes, and nucleation events that initiate failure is therefore critical for engineering resilient dielectric optics. Femtosecond laser pulses deposit energy on ultrafast timescales, generating transient extreme pressure (up to 125-330 GPa), temperature ($>$4000 K) conditions that overlap with those found in planetary interiors and warm dense matter regimes \cite{Bazhanova2017High-pressureCore,Knittle1991ModelsEarth}. Such conditions enable access to high-pressure silica polymorphs within confined volumes, without large-scale static compression. Here, we harness femtosecond excitation to drive structural transformations in amorphous silica and probe the resulting states with nanometre-scale resolution. Extensive studies have been carried out recently to understand this structural transformation through molecular dynamics (MD) simulations \cite{ThuyTrang2020StudyCompression,Nhan2019CrystallisationSimulation,Nguyen2024CrystallisationSimulation}. One approach shows that silica can be crystallized isochorically at temperatures higher than the melting temperature of silica glass \cite{Saika-Voivod2006TestSilica}. Such MD pathway can mimic the laser irradiation effect inside a bulk silica glass, where the nonlinear absorption of ultrashort laser irradiation affects small volumes inside the bulk \cite{doi:10.34133/ultrafastscience.0056}. The temperature evolution can be considered isochoric due to the constraints of the surrounding cold environment \cite{thermodynamicPathways2025}. However, the study relied on the Berendsen thermostat to quench the liquid to the target temperature, which, although it offers numerical stability and simplicity, does not generate a true canonical ensemble. Instead, it rescales velocities each timestep to enforce a target temperature with the damping, suppressing natural temperature fluctuations that are essential for correctly sampling thermodynamic states. In the context of laser-induced melting and recrystallization, this is particularly important, since highly localized and nonequilibrium thermal processes dominate in this regime. This will in the end affect the atomic positions of the system before the equilibration, which will in turn have a substantial impact on the Gibbs free energy calculations and, as a consequence, the speed of the nucleation. As our study concentrates on recreating laser-driven crystallization of silica, we adopt a more physically justified ensemble, discussed in the computational details section.

In this study, we investigate the femtosecond-laser-induced formation of high-pressure phases in amorphous silica through a combined experimental and computational approach. Transmission electron microscopy (TEM) and four-dimensional scanning transmission electron microscopy (4D-STEM) provide high-resolution structural and diffraction imaging of laser-irradiated regions, revealing nanoscale motifs that deviate from the ambient amorphous network. These experimental observations are complemented by molecular dynamics (MD) simulations that model the rapid heating, pressurization, and quenching behavior inherent to femtosecond laser interaction. In our MD simulations, we studied the synthesis of stishovite phase from amorphous silica and phase transition from stishovite to pyrite phase. We use several snapshots from the simulation to calculate Gibbs free energy by Density Functional Theory (DFT) to study the crystallization process at non-ambient conditions. We then adapt the simulation parameters to capture phase transition of stishovite to pyrite crystal. The thermodynamic conditions of such crystallization and phase transition regimes will complement our experimental results and create a theoretical network for these processes. Together, these tools allow us to map coordination transitions and identify signatures of metastable or partially crystalline high-pressure phases. Our findings provide direct evidence of laser-induced structural transformations in silica and contribute to a deeper understanding of non-equilibrium phase behavior in dielectric materials under extreme conditions.

\section{Results}\label{sec2}

In our earlier experimental work \cite{Noor2025PulseSystems}, femtosecond laser irradiation of SiO$_2$/HfO$_2$ multilayer dielectric (MLD) mirrors under single-shot excitation at 1030 nm was shown to produce well-defined subsurface blisters arising from ultrafast electronic excitation, confined energy deposition, and stress accumulation within the top silica layers. Using pulse durations of 260 fs, 77 fs, and 25 fs, we measured single-shot laser-induced damage thresholds (LIDT) of 2.29 J/cm$^2$, 1.63 J/cm$^2$, and 0.98 J/cm$^2$, respectively. The onset of damage manifested as circular blisters whose diameter and height scaled systematically with local laser fluence and pulse duration. Atomic force microscopy (AFM) revealed Gaussian blister height profiles reflecting the spatial fluence distribution of the incident beam. Cross-sectional transmission electron microscopy (TEM) and finite-difference time-domain (FDTD) simulations further confirmed that blister formation was confined to the upper SiO$_2$/HfO$_2$ layers, where localized stress accumulation driven by electron–hole plasma expansion and subsequent lattice heating led to buckling and interfacial delamination. These studies established the fluence- and pulse-duration-dependent blister morphology and identified delamination as a dominant precursor to macroscopic damage.

Building on this experimentally established damage framework, the present work goes beyond morphological characterization to directly resolve the nanoscale structural state of the laser-modified amorphous silica within these confined regions. Using cross-sectional TEM, selected-area electron diffraction (SAED), and four-dimensional scanning transmission electron microscopy (4D-STEM), we demonstrate that the stress-confined blister volume hosts localized crystallization into multiple silica polymorphs, including high-pressure phases not accessible under equilibrium conditions. These observations establish a direct link between ultrafast stress confinement, subsurface blister formation, and pressure-driven structural transformation in amorphous silica, providing a structural foundation for interpreting femtosecond laser-induced damage pathways. 

\subsection{Blisters Formation}\label{subsec1}

\begin{figure}[!htbp]
\centering
\includegraphics[width=1\textwidth]{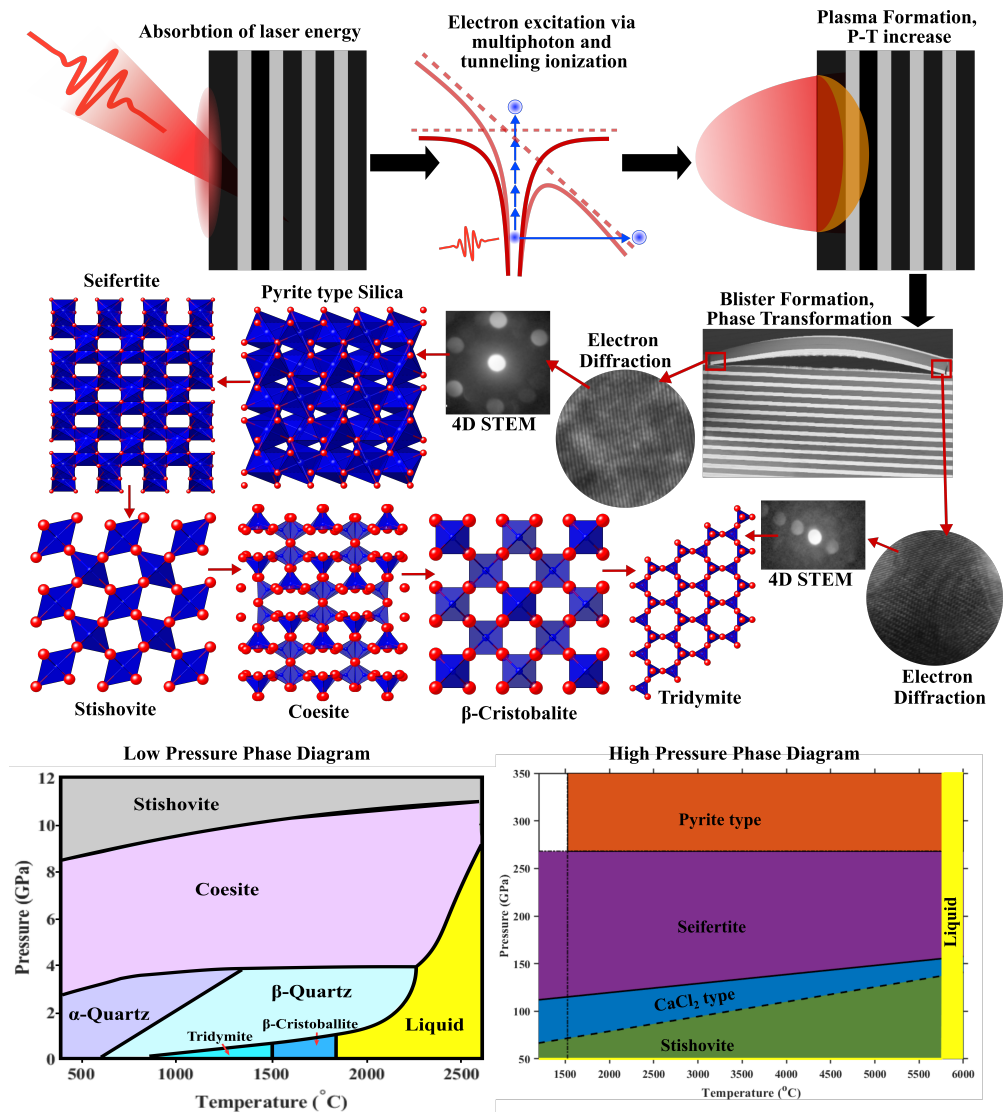}
\caption{Plasma-driven blister formation and pressure-induced phase transformation in silica under ultrafast laser excitation: Schematic illustration of femtosecond laser interaction with the SiO$_2$/HfO$_2$ multilayer stack. At high intensity, multiphoton and tunnelling ionization generate a dense electron plasma, leading to rapid energy deposition and bond breaking before lattice relaxation. Subsequent electron–phonon coupling drives ultrafast heating and volumetric expansion, producing subsurface void nucleation and blister formation. Confinement of the expanding material within the multilayer stack generates transient extreme pressures and temperatures, enabling stabilization of metastable high-density silica polymorphs upon rapid quenching. For reference, the lower panel shows representative low-pressure \cite{https://doi.org/10.1029/93JB02968,Yagi1976DirectMeasurements,Precise1982CoesiteComponents}and high-pressure silica phases \cite{Kuwayama2005,Andrault2014PhaseConditions,Oganov2005,Fischer2018EquationsSiO2}.Phase diagrams are computed in matlab based on literature values and edited in Inkscape vector graphics.}
\label{fig1}
\end{figure}

Blister formation in multilayer dielectric (MLD) mirrors under femtosecond laser irradiation is generally understood to arise from ultrafast electronic excitation, nonlinear absorption, and subsequent stress confinement within the layered structure. Previous studies have shown that dense electron–hole plasma generation and rapid energy transfer to the lattice produce localized thermomechanical stress, which can exceed the interfacial adhesion strength and lead to delamination without immediate material removal \cite{Talisa2020,Zhang2022d,Chen2014}. In such systems, the low thermal diffusivity of the constituent oxides and the multilayer geometry play a critical role in trapping both energy and stress on ultrafast timescales \cite{Talisa2019b}. Here, rather than focusing on blister morphology or failure thresholds, we directly examine the structural consequences of stress-confined blister formation within the silica layers of the MLD stack. Cross-sectional electron microscopy of laser-generated blisters reveals that the mechanically confined region beneath the blister cap is not merely elastically or plastically deformed, but undergoes substantial local densification and structural rearrangement. These confined subsurface volumes therefore constitute localized environments in which extreme, transient pressure–temperature conditions can develop during femtosecond laser excitation. Figure \ref{fig1} schematically summarizes this process: nonlinear absorption in the region near the interface of silica-hafnia layers generates a dense electron–hole plasma excited by the intensification of the laser field created by the MLD structure, producing an initial electronic pressure surge on sub-picosecond timescales. Subsequent lattice heating and volumetric expansion lead to stress accumulation at the SiO$_2$/HfO$_2$ interface, driving buckling and delamination that form a subsurface blister \cite{Noor2025PulseSystems}. In the present work, we demonstrate that this stress-confined blister volume serves as the site of laser-driven structural transformation in amorphous silica.

The structural evolution of silica under varying thermodynamic conditions has been extensively characterized under equilibrium and near-equilibrium compression pathways \cite{Keskar1992,Koike2013InfraredFormation,Dove2008KineticsPolymorphs,Hu2015PolymorphicCoesite,Sato2010High-pressureGPa}. At ambient pressure, silica adopts tetrahedrally coordinated polymorphs such as $\alpha$-quartz, tridymite, and $\beta$-cristobalite, which are related through displacive and reversible transformations along the temperature axis. Under increasing pressure, however, silica undergoes reconstructive phase transitions marked by a change in silicon coordination. Above approximately 9 GPa, tetrahedrally coordinated networks collapse into the octahedrally coordinated stishovite phase, accompanied by a large density increase from $\sim$2.2 g cm$^{-3}$ to $\sim$4.3 g cm$^{-3}$. At still higher pressures, stishovite can transform into denser polymorphs such as seifertite and, at ultrahigh pressures, the pyrite-type structure \cite{karki1997abinitio, teter1998highPressure}, both of which have been reported in static high-pressure and shock-compression experiments. In the context of the present experiments, these established phase relations provide a reference framework for interpreting the crystallographic signatures observed within laser-generated blisters. Unlike static compression, femtosecond laser irradiation applies pressure and temperature on ultrafast timescales and within nanometer-scale confined volumes. As shown in the following sections,  electron diffraction and 4D-STEM measurements reveal that the blister-confined silica hosts localized crystalline domains corresponding to both high-pressure and intermediate-pressure polymorphs. The coexistence of these phases within a single laser-modified region indicates that femtosecond laser excitation drives silica along a non-equilibrium thermodynamic pathway that differs fundamentally from conventional static or thermal processing. The implication of this finding is tremendous, as unlike in the case of DAC experiments with CW laser heating, these metastable phases inside the multilayer stack appear to be in a long lived metastable condition so that they can be studied long after the laser pulse creates them. Therefore, in future, other materials of interest (e.g. Fe, Mg, Fe-O, C) can be inserted into these dielectric stacks to observe formation of high pressure phases while interacting/reacting with SiO$_2$ or oxygen. Some of these interactions appear to be very active in the core-mantle boundaries. 

\subsection{Ultrafast Laser-Induced Damage and Electron Diffraction}\label{subsec2}

\begin{figure}[!htbp]
\centering
\includegraphics[width=1\textwidth]{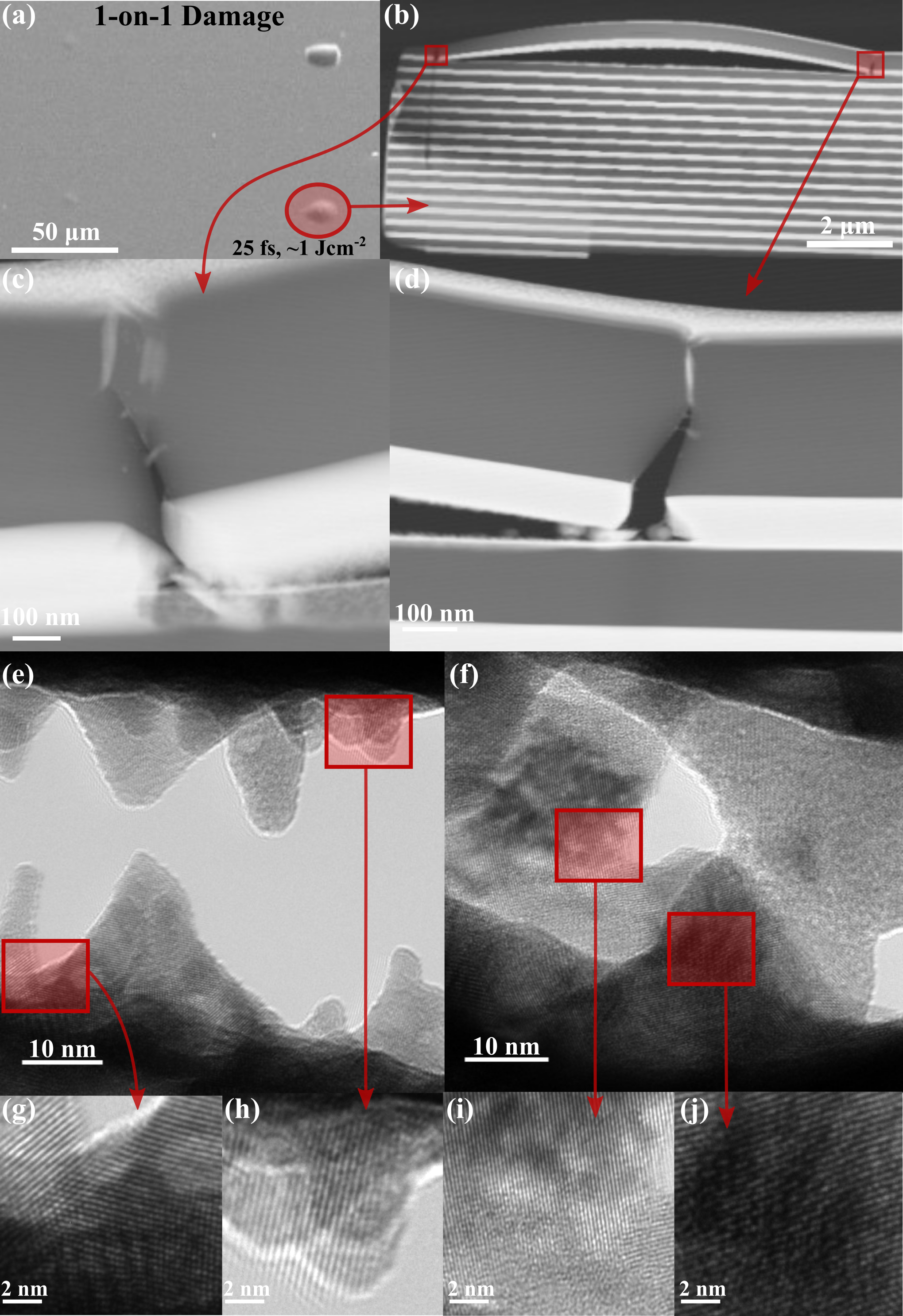}
\caption{(a) SEM image of blister generated inside amorphous SiO2/HfO2 MLD structure interacting with a single, 25 fs ultrafast laser pulse at 1030 nm. The red shaded enclosure is drawn to identify the blistered region used for this study. (b) is FIB cutout of the blister (STEM images) (c) and (d) are the close-up view of nano cracks viewed through STEM. (e), (f) are the HRTEM image of are of used from electron diffraction whereas the (g), (h), (i) and (j) are closer view of the enclosures which show clear crystalline phases.}
\label{fig2}
\end{figure}

To investigate the structural modifications induced by ultrafast laser excitation, we employed a combination of focused ion beam (FIB) milling and transmission electron microscopy (TEM). Site-specific cross-sectional lamellae were prepared from laser-irradiated regions using a dual-beam FIB-SEM system. A protective platinum capping layer was first deposited on the region of interest to prevent surface damage during milling. Subsequent ion milling was performed in stages to achieve a final thickness of $\sim$100 nm, sufficient for electron transparency and high-resolution imaging. The thinned lamella was then transferred to a copper TEM grid for structural analysis. TEM imaging was conducted using both bright-field and high-resolution modes to investigate the morphology and atomic structure of the laser-modified zones. In addition to direct imaging, selected area electron diffraction (SAED) was performed to identify the crystalline and phase composition of the transformed regions. High-resolution TEM (HRTEM) was used to resolve lattice fringes, enabling the identification of nanocrystalline phases that formed under localized high-pressure and high-temperature conditions. Figure \ref{fig2} illustrates the multiscale imaging workflow used to assess the laser-induced structural changes in silica. Fig. \ref{fig2}(a) shows a top-down SEM image of the surface, where multiple blisters are visible because of femtosecond laser exposure. The circled region marks the specific blister selected for further cross-sectional analysis. Fig. \ref{fig2}(b), a FIB-prepared cross-section through the blister cap reveals the internal structure of the modified region, including buckling, subsequent resulting delamination and void formation beneath the surface. Fig. \ref{fig2}(c)$\&$(d) provide close-up STEM views of the cross-section, highlighting the presence of nano-cracks and interfacial separations. These features suggest localized mechanical failure, likely driven by shock wave propagation and rapid volume expansion during energy deposition. Fig. \ref{fig2}(e)$\&$(f) present TEM images of the interior of the void and the surrounding compressed shell. The high contrast features visible in these regions indicate a transition from amorphous to more ordered structures. The red dashed boxes identify regions selected for further high-resolution analysis. Fig. \ref{fig2} (g), (h), (i), $\&$ (j) display HRTEM images from the boxed regions in (e) and (f), revealing distinct lattice fringes with periodic contrast. These fringes confirm the formation of nanocrystalline polymorphs of silica, indicating that the transient P–T conditions generated during the femtosecond laser pulse were sufficient to drive localized phase transformations. The sharp lattice ordering and confined crystallization zones are consistent with rapid quenching from a high-energy state. 

\begin{figure}[!htbp]
\centering
\includegraphics[width=1\textwidth]{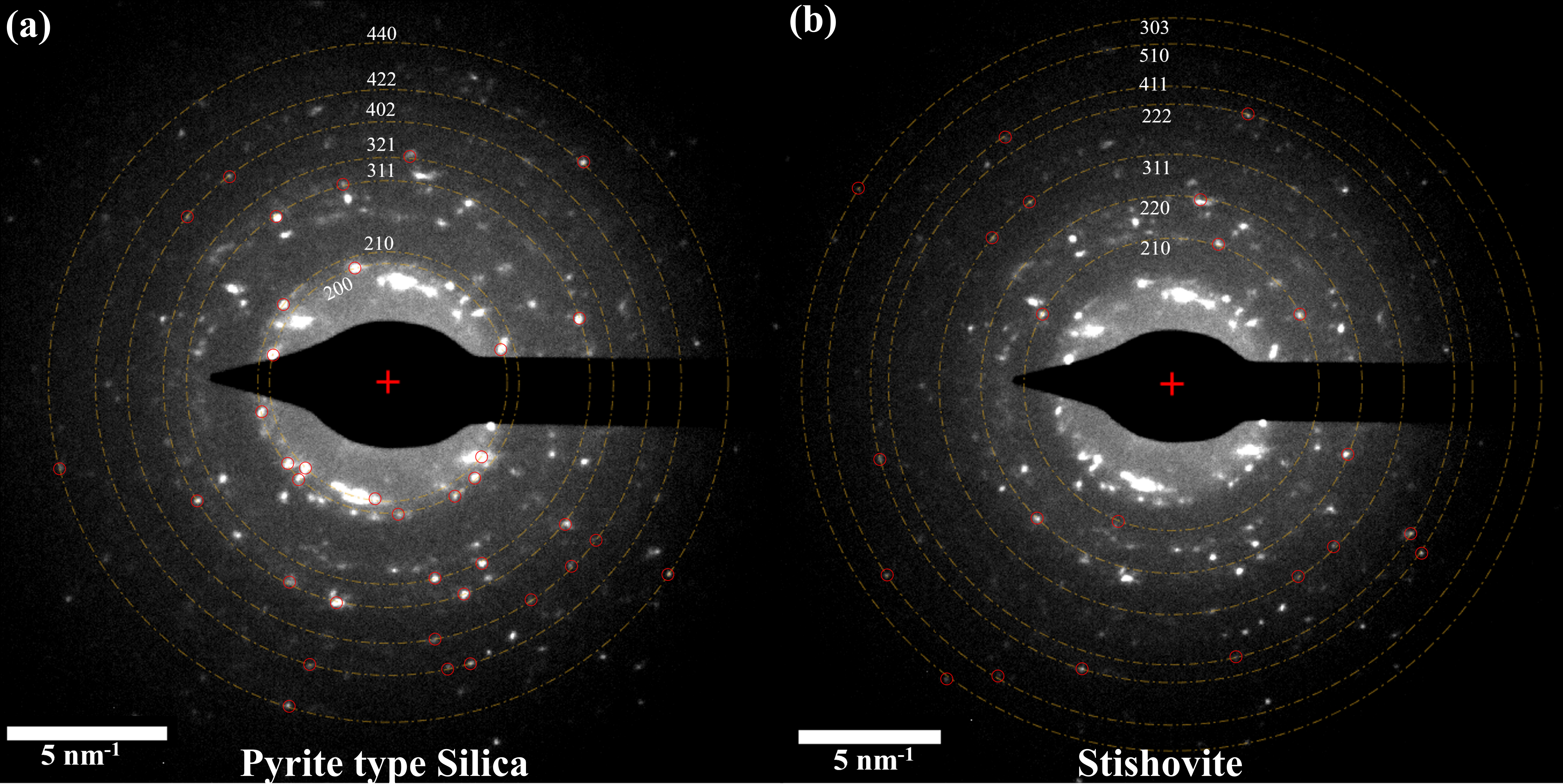}
\caption{Visual representation of high-pressure phases where the red circle are diffraction spots with matches theory and yellow circle is the representation of hkl planes}\label{fig3}
\end{figure}

 High-resolution TEM images of the laser-modified regions reveal the presence of localized crystallinity embedded within an otherwise amorphous silica matrix. To determine the structural identity of these nanocrystalline domains, we performed selected-area electron diffraction (SAED) at multiple locations within laser-irradiated regions. Figure \ref{fig3} shows representative SAED patterns acquired from two distinct blister-confined volumes, highlighting diffraction signatures consistent with high-pressure silica polymorphs. The pattern in Fig. \ref{fig3}(a) is indexed to the pyrite-type structure, while Fig. \ref{fig3}(b) corresponds to seifertite, both of which are high-density phases stabilized under extreme pressure conditions. SAED patterns were recorded using a selected-area aperture to isolate regions with diameters below $\simeq$ 500 nm, ensuring localized structural analysis of individual nanocrystalline domains. Interplanar spacings (\textit{d}-spacings) were quantitatively extracted from the diffraction patterns using Gatan DigitalMicrograph. To identify the corresponding crystalline phases, structural models of SiO$_2$ polymorphs were generated using CrystalMaker, and their theoretical reflection lists were calculated assuming powder-averaged diffraction, consistent with the nanoscale and randomly oriented nature of the crystallites.

\begin{table}[!htbp]
\centering
\caption{Summary of measured and matched diffraction peaks.}
\begin{tabular}{cccccc}
\toprule
\textit{d\textsubscript{measured}} (Å) & \textit{{d\textsubscript{hkl}}} (Å) & ${\Delta d}$ (\%) & \textit{hkl} & {Intensity} & {Matched Polymorph} \\
\midrule
1.952 & 1.9649 & 0.659 & 200   & Moderate & Pyrite-Type \\
1.398 & 1.389  & 0.616 & 220   & Strong   & Pyrite-Type \\
1.177 & 1.184  & 0.667 & 311   & Moderate & Pyrite-Type \\
1.457 & 1.445  & 0.830 & 321   & Strong   & Pyrite-Type \\
1.100 & 1.103  & 0.320 & 422   & Moderate & Pyrite-Type \\
1.008 & 1.004  & 0.398 & 423   & Moderate & Pyrite-Type \\
1.570 & 1.556  & 0.897 & 130   & Strong   & Seifertite \\
1.500 & 1.499  & 0.050 & 221   & Strong   & Seifertite \\
1.268 & 1.264  & 0.240 & 311   & Moderate & Seifertite \\
1.514 & 1.513  & 0.005 & 202   & Moderate & Seifertite \\
1.280 & 1.288  & 0.637 & 23    & Moderate & Seifertite \\
0.951 & 0.947  & 0.434 & 411   & Strong   & Stishovite \\
0.995 & 0.989  & 0.536 & 222   & Moderate & Stishovite \\
0.933 & 0.938  & 0.588 & 312   & Moderate & Stishovite \\
1.129 & 1.124  & 0.434 & 202   & Strong   & Stishovite \\
0.788 & 0.792  & 0.463 & 332   & Weak     & Stishovite \\
2.709 & 2.694  & 0.554 & 131   & Strong   & Coesite \\
1.659 & 1.655  & 0.210 & -423  & Strong   & Coesite \\
3.128 & 3.103  & 0.796 & 20-2  & Strong   & Coesite \\
1.842 & 1.837  & 0.238 & 330   & Strong   & Coesite \\
1.707 & 1.713  & 0.353 & 260   & Moderate & Coesite \\
1.635 & 1.646  & 0.695 & 210   & Strong   & Tridymite \\
1.539 & 1.538  & 0.057 & 105   & Strong   & Tridymite \\
0.847 & 0.853  & 0.756 & 325   & Moderate & Tridymite \\
2.495 & 2.515  & 0.795 & 110   & Strong   & Tridymite \\
1.842 & 1.842  & 0.010 & 112   & Strong   & $\beta$ Quartz \\
1.814 & 1.814  & 0 & 220   & Strong   & Cubic Hafnia \\
2.567 & 2.565  & 0.43 & 200   & Moderate   & Cubic Hafnia \\
0.9866 & 0.987  & 0.04 & 311   & Weak   & Cubic Hafnia \\
2.942 & 2.932  & 0.34 & 101   & Strong   & Tetragonal Hafnia \\
1.518 & 1.522  & 0.010 & 211   & Moderate   & Tetragonal Hafnia \\
1.441 & 1.441  & 0 & 032   & Strong   & Monoclinic Hafnia \\
1.085 & 1.0905  & 0.5 & 432   & Strong   & Monoclinic Hafnia \\
\bottomrule
\end{tabular}
\end{table}
Phase identification was carried out by directly comparing experimentally measured \textit{d}-spacings with simulated values using a custom MATLAB routine that overlays theoretical diffraction rings onto the experimental patterns. Only reflections with deviations below 1 $\%$ were retained and indexed with their corresponding Miller indices (hkl). The resulting agreement between experimental and simulated diffraction features enables unambiguous assignment of both the polymorph and its crystallographic planes, as highlighted by the red-circled reflections in Fig. \ref{fig3}. Analysis of multiple SAED patterns acquired from different laser-modified regions consistently reveals nanocrystalline high-pressure silica phases, confirming that femtosecond laser excitation drives a transformation from amorphous silica to crystalline polymorphs such as pyrite-type silica and seifertite within confined subsurface volumes. The extracted d-spacings from representative selected-area diffraction patterns are summarized in Table 3 and show excellent agreement with theoretical values across multiple sites, demonstrating the reproducibility of the observed phase transformations.

\subsection{4D STEM Analysis}\label{subsec3}

\begin{figure}[h!]
\centering
\includegraphics[width=1\textwidth]{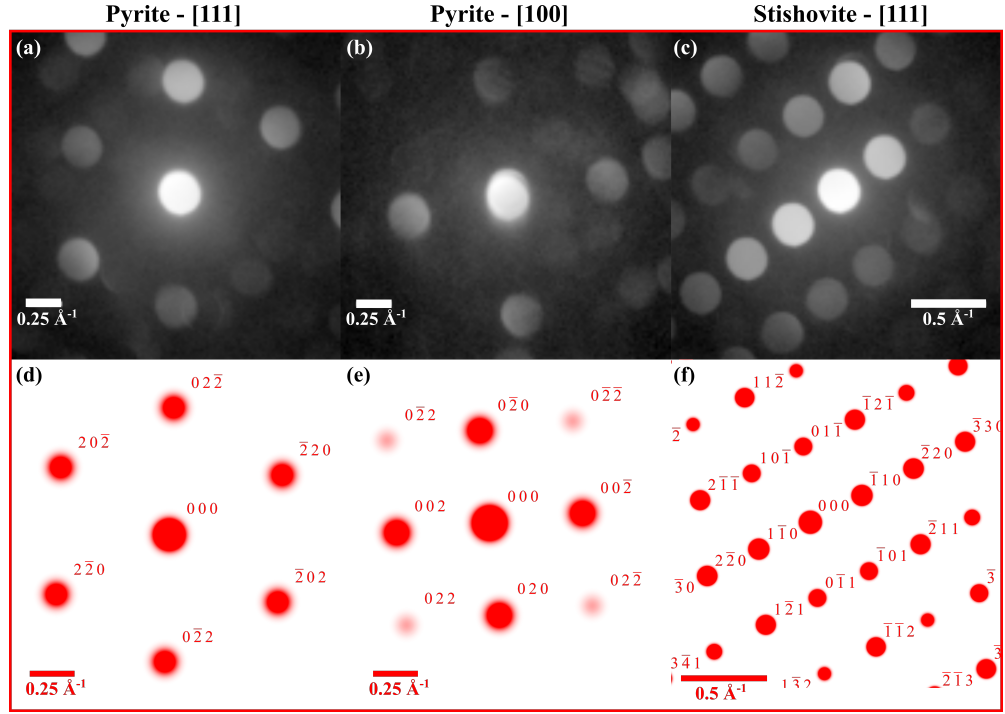}
\caption{(a), (b), (c) are the 4D STEM experimental data obtained at electron diffraction of nanocrystalline regions. (e), (f) is the matched pyrite type silica simulated in single crystal diffract software at [111] and [100] direction. (g) is the simulated stishovite at [111] direction. The experimental data aligns with these simulations with a deviation of less than 5$\%$}\label{fig4}
\end{figure}

To determine the crystallographic identity of the laser-induced phases with nanoscale resolution, we employed four-dimensional scanning transmission electron microscopy (4D-STEM), which records a full diffraction pattern at each probe position and enables spatially resolved mapping of reciprocal-space signatures from individual nanodomains embedded within the surrounding amorphous matrix. This approach allows phase identification without averaging over heterogeneous regions. The experimental diffraction patterns (Fig. \ref{fig4}a–c) display sharp, symmetry-defined reflections indicative of long-range order. Direct comparison with simulated single-crystal diffraction patterns (Fig. \ref{fig4}d–f) reveals excellent agreement with pyrite-type silica along the [111] and [100] zone axes and stishovite along [111]. Extracted reciprocal-lattice spacings match simulated values within $\sim$5$\%$, providing quantitative confirmation of phase identity. The spatial confinement of these high-density polymorphs within the laser-modified interface, together with their coexistence with lower-density phases such as tridymite, indicates a strongly non-equilibrium transformation pathway. These observations are consistent with transient extreme pressures and temperatures followed by ultrafast quenching, stabilizing multiple metastable silica polymorphs within nanometre-scale volumes.

\subsection{Molecular Dynamics Simulations}\label{subsec4}

\begin{figure}[!htbp]
\centering
\includegraphics[width=1\textwidth]{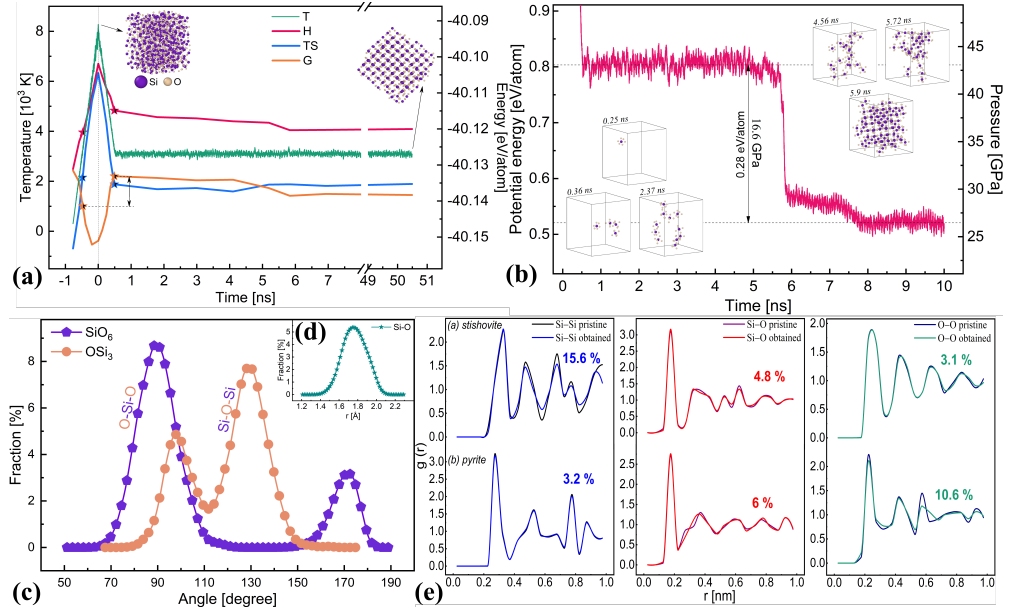}
\caption{(a) Temporal profile of temperature during the amorphization and crystallization of stishovite (green) and evolution of Gibbs free energy (G) and its components, enthalpy (H) and entropy (S) (shifted by a value of -40.15), calculated for several snapshots over the MD timescale by DFT simulations. Time is shifted so that the cooling starts at 0 ns. (b) Temporal evolution of the per-atom potential energy and the pressure during the first 10 ns of the equilibration. The potential energy is averaged over the whole system including the Si and O atoms. (c) Bond-angle distribution for O-Si-O and Si-O-Si angles in SiO$_6$ and OSi$_3$ units, inset (d) shows the distribution of Si-O bonds. (e) Partial RDFs of Si-Si, Si-O and O-O pairs of the obtained (a) stishovite and (b) pyrite structures compared with pristine crystals at the same density.}\label{fig5}
\end{figure}

To investigate the crystallization behavior of fused silica under extreme thermodynamic conditions, large-scale molecular dynamics (MD) simulations were performed using a modified BKS potential. The system, consisting of a stishovite supercell, was thermally cycled to induce melting and subsequent recrystallization. After equilibration at 300 K, the structure was heated to 8000 K to melt and remove crystalline memory, followed by rapid quenching at $10^{13}$~K/s. The system was then held at different temperatures to probe nucleation, with crystallization occurring at 3100 K. Figure \ref{fig5}(a) shows the thermodynamic evolution. Gibbs free energy, calculated from DFT snapshots, and its enthalpic and entropic components reveal the driving mechanisms. During heating, increased internal energy and entropy reduced the free energy, leading to melting. Upon cooling, the system entered a supercooled liquid state with gradual free energy decrease. At approximately 5.5 ns, sharp changes in energy and pressure (Fig. \ref{fig5}(b)) marked spontaneous crystallization. A rapid drop of $\sim0.28$ eV/atom in potential energy and $\sim16.6$ GPa in pressure within 0.5 ns indicates atomic rearrangement into a dense crystalline phase. This behavior is consistent with classical nucleation theory, where the system remains metastable until a critical nucleus forms, followed by fast growth.

Local structural analysis was performed using a Python-based tool to identify SiO$_6$ octahedra based on bond-length and angular criteria. Transient octahedral clusters appeared during cooling ($\sim25$ ns) but dissolved until the nucleation threshold at 5.5 ns. After nucleation, the crystalline phase rapidly propagated. Angle distributions (Fig. \ref{fig5}(c)) show O-Si-O peaks near 90$\textdegree$ and 172$\textdegree$, corresponding to adjacent and trans ligands in ideal octahedra, confirming stishovite-like ordering. Similarly, Si-O-Si angles display bimodal peaks at 98$\textdegree$ and 130$\textdegree$, indicating edge- and corner-sharing octahedra. The Si-O bond length peaks sharply at 1.75 $\AA$, consistent with experimental values (1.73 $\AA$ at 35 GPa), supporting structural realism. Partial radial distribution functions (PRDFs) were compared with pristine stishovite at the same density (Fig. \ref{fig5}(d)). First-neighbor Si-O and O-O peaks show excellent agreement, with RMSD values of 4.8\% and 3.1\%, indicating preserved short-range order. The Si-Si RDF exhibits higher deviation (15.6\%) due to lattice misalignment from randomly oriented nucleation events affecting medium-range correlations. We further examined the transformation to the denser pyrite phase. At 2400 K, uniaxial compression along [110] up to 180 GPa initiated structural changes, while additional compression along [010] to 200 GPa completed the transition. RDF analysis (Fig. \ref{fig5}(e)) confirmed agreement with reference pyrite at equivalent density. The transition proceeds displacively through concerted atomic rearrangements, consistent with direction-dependent transformation mechanisms under extreme stress. Overall, simulations show that fused silica crystallizes into stishovite through a mostly entropy-driven process and can transform into pyrite under anisotropic ultrahigh pressure. These transitions are feasible within extreme P-T-t conditions generated by femtosecond laser irradiation. Combined MD and DFT analysis provides atomic-level insight into silica polymorphism relevant to materials science and planetary physics \cite{Kuwayama2005}.

\section{Methods}
\subsection{Ultrafast Laser Damage and Electron Diffraction Characterization}

Ultrafast laser-induced damage experiments on the MLD high-reflectivity mirror were performed using a commercial PHAROS laser system operating at 1030 nm with 25 fs pulse duration. The incident beam was \textit{p}-polarized at an angle of 45$\degree$. Details of the experimental configuration and damage threshold measurements are described in \cite{Noor2025PulseSystems}. Ultrafast laser–induced structural modifications in amorphous silica were investigated using site-specific cross-sectional transmission electron microscopy (TEM) combined with electron diffraction techniques. Laser-modified regions containing surface blisters were first identified by scanning electron microscopy (SEM). Cross-sectional lamellae were prepared from selected blister sites using a dual-beam focused ion beam (FIB–SEM) system. A protective Pt capping layer was deposited prior to ion milling, and the lamellae were thinned to a final thickness of approximately $\sim$100 nm for electron transparency before transfer to Cu TEM grids. Bright-field TEM, scanning TEM (STEM), and high-resolution TEM (HRTEM) were employed to resolve nanoscale cracking, void formation, and localized crystalline domains embedded within the amorphous matrix. Selected-area electron diffraction (SAED) patterns were acquired using apertures below $\sim$500 nm in diameter to probe individual nanocrystalline regions. Interplanar spacings were extracted using Gatan DigitalMicrograph and compared against simulated diffraction reflections generated from structural models of SiO$_2$ polymorphs using CrystalMaker. Phase identification was performed through direct matching of experimental d-spacings with theoretical values, enabling assignment of high-pressure silica phases including pyrite-type silica, seifertite, and stishovite. Complementary four-dimensional STEM (4D-STEM) was further used to map diffraction contrast with nanoscale spatial resolution by recording a full diffraction pattern at each probe position. Experimental diffraction patterns were compared with simulated zone-axis patterns from SingleCrystal software, confirming phase diversity within confined laser-modified subsurface volumes.

\subsection{Computational Details}
For the simulations modified BKS (Beest-Kramer-Santen) potential \cite{Saika-Voivod2001ComputerTransition} (i.e.. the original BKS potential \cite{Beest1990ForceMODEL} with an additional 30-6 Lennard-Jones-type term \cite{Mie1903ZurKorper}) was used with periodic boundary conditions. BKS potential was extensively used in dozens of studies to describe both liquid and solid states \cite{Karki2007First-principlesPressure,Karki2010Visualization-basedMelt,Hung2007DiffusionSimulation}.

A unit cell of stishovite \cite{Pavese1995X-rayK} with lattice parameters of 4.17 and 2.66 \AA~consisting of 6 atoms was replicated to create a supercell of 1350 atoms. The density of the system was 4.35 g/cm$^3$ with a good agreement of the experimental density of 4.29~g/cm$^3$. Starting with a random distribution of oxygen and silicon atoms might initially create defects which might affect the final outcome of the simulations. Hence, we start our simulations from a stishovite supercell to obtain amorphization at the needed density. We employ the canonical NVT ensemble using a Nosé-Hoover thermostat maintaining proper thermodynamic fluctuations and allowing for a more realistic representation of local cooling behavior in the small, dense volume simulated in our study. At first, it was thermalized at 300 K for 1 ns. Then, the system was heated to 8000 K to ensure complete amorphization and removal of any crystalline memory. The system was cooled to different temperatures afterwards with a rate of 10$^{13}$ K/s and a damping parameter of 10 ps. It was kept at these temperatures for 50 ns, a relatively long time to ensure the possible crystallization and full relaxation. The investigation showed that the system underwent a pure crystallization at 3100 K, which is going to be studied in next sections.

\backmatter
\bmhead{Acknowledgements}
U.S. Department of Energy (STTR DE-SC0019900);  AFOSR award no. FA9550-25-1-0297. We would like to thank Drs. Daniel Veghte, Daniel Huber and Amanda Trout, of the center for electron microscopy and analysis (CEMAS) at the Ohio State University for helping with the SEM/FIB/STEM
images, respectively. A.Y. thanks Ivan Saika-Voivod for the useful discussion.
\section*{Declarations} The Authors declare no conflict of interest.
\bibliography{references}
\end{document}